\documentstyle[eqsecnum,aps,epsf]{revtex} % PH. REV. FINAL FORMAT STYLE
\newcommand{\postscript}[2] {\setlength{\epsfxsize}{#2\hsize}
\centerline{\epsfbox{#1}}}

\begin{document}
\twocolumn[\hsize\textwidth\columnwidth\hsize\csname@twocolumnfalse\endcsname

%\begin{document}

%\begin{center}

\title{\bf Low-energy quenching of positronium by helium}

\author{ Sadhan K. Adhikari,  P. K. Biswas, 
and R. A. Sultanov}
\address{Instituto de F\'{\i}sica Te\'orica, 
Universidade Estadual Paulista,
01.405-900 S\~ao Paulo, S\~ao Paulo, Brazil\\}

%\end{center}
\date{\today}
\maketitle

\begin{abstract}

Very low-energy scattering of orthopositronium by helium has been
investigated for simultaneous study of  elastic cross section
and pick-off quenching rate using a model exchange potential. The present
calculational scheme, while  agrees with the measured cross section of
Skalsey et al,   reproduces successfully  the   parameter
$^ 1Z_{\makebox{eff}}$, the effective number of electrons per atom in a
singlet state relative to  the positron. Together with the fact that this
model potential  also leads to an agreement with measured medium energy
cross sections of this system,  this study  seems to  resolve the
long-standing discrepancy at low energies among  different theoretical
calculations and experimental  measurements.

{\bf PACS Number(s): 34.10.+x,  36.10.Dr} 
\end{abstract} 
\vskip1.5pc]

Studies on positronium- (Ps-)impact scattering have gained momentum these
days due to the availability of ortho-Ps beam in the laboratory and its
vast applicational potential coupled with the present inadequate and
inconclusive understanding of its interaction dynamics with matter
\cite{g1}.  Ps scattering by neutral targets has posed a challenge to
theoreticians on a proper accounting of experimental data as most existing
theoretical works disagree with the major experimental trend. 

The discrepancy figures out prominently in the Ps-He system where
there are many theoretical and experimental studies.  The medium-energy
experimental cross section
shows a declining trend with decreasing energy \cite{l1} from a peak
around 20 eV for Ps-He scattering.  
Similar
trend is also observed in Ps-H$_2$ and Ps-Ar systems \cite{l1,l2}. 
This trend, which is supported by
the recent measurement of Skalsey et al \cite{g2}, could not be reproduced
in most theoretical predictions \cite{d2,h,bb,gh2,fk1}.  
At very low energies, these theories and experiments \cite{l1,l2,g2,n,n1}
on the Ps-He system are inconsistent with each other and also among
themselves.  For illustration, the zero-energy cross sections calculated
on Ps-He
by different authors vary from $3.3 $\AA$^2$
\cite{gp} to $16.54$ \AA$^2$ \cite{fk} while the measured values range
from
$2.3 \pm 0.4$ \AA$^2$ (at 0.915 eV) \cite{g2} to $11 \pm 3$ \AA$^2$ (between
0 to 0.3 eV) \cite{n}.  Pointing out the very reactive 
nature of Ps scattering and its associated convergence difficulties, a
prescription for the generation of nonlocal model exchange potential has been
advocated recently
 and applied successfully to different electron-impact
(targets: H, He) and Ps-impact (targets: H \cite{ba1,ba1a}, He \cite{ba2},
H$_2$
\cite{ba3}, Ar, Ne \cite{ba4}) scatterings
problems
 using static-exchange to three-Ps-state models.
The three-Ps-state calculation for Ps-He
 predict a further lower  zero-energy cross section of
 2.42 \AA$^2$ \cite{ba2}.

In this work we shed light on the abovementioned discrepancy in Ps-He
system in conjunction with a determination of the parameter
$^1Z_{\makebox{eff}}$ which denotes the effective number of electrons per
atom in a singlet state relative to the positron.  The study of this
parameter is supposed to provide a stringent test for the model potential.
The incident ortho Ps(1s) atom in a triplet state with a lifetime of 142ns
can decay into three photons and is more stable than its para counterpart
in a singlet
state with a lifetime of 0.125ns for a two-photon decay mode.  However, in
its interaction with matter, the positron of Ps can find an atomic
electron in a spin-singlet state and the two can be annihilated by a 
two-photon decay mode without really forming a para Ps atom by electron
exchange.  This process is termed pick-off quenching.  From the
experimental pick-off quenching rate the parameter $^1Z_{\makebox{eff}}$
can be extracted. Theoretically $^1Z_{\makebox{eff}}$ can be calculated
from the wave function of the Ps-He system $\Psi({\bf r_p},s_p;{\bf
r_1},s_1; {\bf r_2}, s_2;{\bf r_3},s_3)$ where ${\bf r}$ and $s$ refer to
position and spin, the suffix $p$ refers to the positron and $i=1,2,3$
refer to the electrons.  Following Barker and Bransden \cite{bb,fk}, the
amplitude for finding the positron and one of the atomic electrons in a
relative singlet state is 
\begin{equation}\label{1} \Phi({\bf r_p;
r_1;r_2,}s_2; {\bf r_3}, s_3)=< \chi_0(s_p,s_1)  |\Psi>, \end{equation}
where $\chi_0$ is the singlet wave function. The parameter
$^1Z_{\makebox{eff}}$ is given by \begin{equation}\label{2} ^1
Z_{\makebox{eff}}=3\sum_ {\makebox{spin}} \int d {\bf r_p} d {\bf r_1} d
{\bf r_2} d {\bf r_3} \delta ({\bf r_p-r_1}) |\Phi|^ 2.  \end{equation}
The factor 3 appears as each of the three electrons of the Ps-He system
contributes equally to $^1Z_{\makebox{eff}}$. 

Unlike the scattering cross sections, which are determined from the 
asymptotic part of the Ps-He wave function, 
the parameter $^1Z_{\makebox{eff}}$ is sensitive 
to the Ps-He wave function at short 
distances and its correct evaluation in a theoretical calculation 
 should provide 
a sensitive test about its realistic nature. 
There is 
considerable 
discrepancy between theory and experiment in the value of the parameter 
$^1Z_{\makebox{eff}}$. The experimental measurements have yielded 
$^1Z_{\makebox{eff}} =0.108\pm 0.01$ \cite{fk,dh},
$^1Z_{\makebox{eff}} = 0.135\pm 0.068$ \cite{bb,dh1}, 
and $^1Z_{\makebox{eff}} =0.25\pm 25\%$ \cite{bb,rk}, whereas different
static-exchange calculations have yielded values ranging from 0.02 to
0.1  \cite{d2,bb,fk}. Compared to other exchange
potentials, the present scheme
 leads to substantially weaker repulsive exchange
potential and it is expected that the present scheme will lead to
a larger value of $^1Z_{\makebox{eff}}$
 in Ps-He system as is demanded by
experiments \cite{dh,dh1,rk}.

In the static-exchange approximation the Ps-He wave function is represented
by the following antisymmetrized product of 
the internal wave functions of Ps(1s), $\phi_{Ps}({\bf r_1-r_p})$, 
and singlet He(1s1s), $\phi_{He}({\bf r_2,r_3})$, 
with a wave function of relative motion, $F_{\bf k}({\bf R})$, 
and a suitable spin function:
\begin{equation} \label{3}
\Psi={\cal A}\phi_{Ps}({\bf t}) \phi_{He}({\bf r_2,r_3})
F_{\bf k}({\bf R})\chi(s_1,s_2,s_3,s_p),
\end{equation}
where ${\bf R=(r_1+r_p)}/2$, ${\bf t=r_1 -r_p}$,
${\cal A}$ is the antisymmetrizer, ${\bf k}$ the incident Ps momentum,
and $\chi$ the spin function.

On expanding $F({\bf R})$ in partial waves
\begin{equation} \label{4}
F_{\bf k}({\bf R})=\sum_{L=0}^\infty (2L+1) (kR)^ {-1}F_L(R)P_L(\cos \theta),
\end{equation}
where $\theta$ is the angle between ${\bf k}$ and ${\bf R}$, the following 
integro-differential equation is obtained from the Schr\"odinger equation:
\begin{eqnarray}\label{5}
\left(\frac{d^ 2}  {dR^ 2}+k^ 2-\frac{L(L+1)}{R^ 2}    \right)F_L(R)\nonumber
\\
=\int_0^\infty  V_L(R,R')F_L(R') R'^ 2  dR',
\end{eqnarray}
where we use 
$V_L(R,R')$ is the nonlocal exchange potential. The asymptotic boundary
conditions for $F_{\bf k}({\bf R})$  and $F_L(R)$ are given by
\begin{eqnarray}
F_{\bf k}({\bf R}) &\sim_{R\to \infty}& \exp(ikR\cos \theta)
+f(\theta) \frac{\exp(ikR)}{R},\\
F_L(R) & \sim_{R\to \infty}& \sin(kR-L\pi/2 +\delta_L),
\end{eqnarray}
where $\delta_L$ is the scattering phase shift, and the
scattering amplitude $f(\theta)$ is given by
\begin{equation}
f(\theta) = \sum_{L=0}^\infty (2L+1)\frac{\exp(i\delta_L)\sin
\delta_L}{k}P_L(\cos \theta).
\end{equation}
The total elastic and momentum transfer cross sections are given by
\begin{eqnarray}
\sigma_{el}(k^ 2)&=& \int |f(\theta)|^ 2 d\Omega,\\
\sigma_m(k^2) &=& \int |f(\theta)|^ 2 (1-\cos\theta)d\Omega,
\end{eqnarray}
respectively. Some of the experiments provide only low-energy
momentum-transfer cross section and hence we also calculate this observable
in this study.

We employ
He(1s1s) wave function of the following form
\begin{eqnarray} \label{6}
\phi_{He}({\bf r_2, r_3})& = & u_{2}({\bf r_2})u_{3}({\bf r_3}) \\
u_{i}({\bf r}) & = & \sum_\kappa
   a_{\kappa i}   \phi_{\kappa i} ({\bf r}
 ) ,\label{6a}
\end{eqnarray}
with $ \phi_{\kappa i} ({\bf r
 }) =\exp (
{-\alpha_{\kappa i} r} ) Y_{00}.$ In the present calculation we use an
accurate
two-term parametrization of (\ref{6a}) \cite{bj}.
In momentum space 
the  model exchange potential  has the following form \cite{ba2}:
\begin{eqnarray}\label{7}
B({\bf k_f,k_i})&=& \biggr[
\sum_j \sum_{\kappa \kappa'}
\frac{4a_{\kappa j} a_{\kappa' j}}{D_{\kappa\kappa 'j}} \nonumber
\\ &\times&
\int \phi_{\kappa'j}^*({\bf r}) \exp ( \makebox{i} {\bf Q. r})\phi_{\kappa j}
({\bf r}) \makebox{d}{\bf r}\biggr]
\nonumber \\ &\times& \int \phi^*_{Ps}({\bf  t })\exp ( \makebox{i}{\bf
Q}.{\bf t }/2)\phi_ {Ps}({\bf  t }) \makebox{d}{\bf  t },
\end{eqnarray}
with  ${\bf  Q = k_i-k_f}$,  $V({\bf p, q})=-B({\bf p,q})/(2\pi^2)$ and
\begin{equation}\label{8}
D_{\kappa\kappa '   j}= [{k_f^2/4+\alpha_{\kappa j}^2+
\beta ^2}]\label{a}
\end{equation}
where 
 $\phi_{\kappa j} ({\bf r}) $ is the $\kappa$th function of the $j$th
electron for the atomic ground state, and  $\phi_{Ps}({\bf t}) = 
\beta^{3/2}\exp(-\beta t)/\sqrt \pi$.  The direct potential for this
problem is zero, and there is a change of sign in the spin-triplet Ps-He
potential below.  The partial-wave configuration space 
nonlocal potential of Eq. (\ref{5}) is given by 
\begin{eqnarray}
V_L(R,R')&=&\left( \frac{2}{\pi}\right)^ 2\int_0^\infty
\int_0^\infty  p^ 2 dp  q^ 2 dq \nonumber \\
&\times& j_L(pR)V_L(p,q) j_L(qR')      \label{9}
\end{eqnarray}
and 
\begin{equation}\label{10}
V_L({p,q})= -\pi^2 \int_{-1}^ 1dx P_L(x)V({\bf p,q}),
\end{equation}
where $x$ is the angle between ${\bf p}$  and ${\bf q}$.
Although the model exchange potential (\ref{10})   has been 
found to be satisfactory for calculating scattering cross sections
\cite{ba2}, it is
interesting to investigate if it is also effective in the calculation of 
finer scattering observables, such as the parameter $^ 1 Z_{\makebox{eff}}.$
A scheme for the calculation of $^ 1 Z_{\makebox{eff}}$ for He wave functions of type 
(\ref{6}) is given by Fraser and Kraidy \cite{fk}
 and we employ the same  in the present calculation.

First we show our results for the low-energy elastic and momentum-transfer
cross sections  in Fig. 1 together those obtained from other theories and
experiments. The triplet Ps-He scattering length in this case is
0.87$a_0$, compared to 1.39$a_0$ obtained by Drachman and
Houston
\cite{d2}.
The discrepancy among various results is apparent in this plot.
The three experimental results for cross sections \cite{g2,n}
shown by solid
circles vary from 2.3 \AA$^2$ \cite{g2}, through 7.45 \AA$^2$
\cite{n1}, to
11 \AA$^2$ \cite{n} at 0.9 eV, 0 eV, and 0.15 eV, respectively.
It is
difficult to reconcile these three experimental
results in a theoretical model. Previous
static-exchange calculations \cite{bb,gh2,fk1,fk} except those of Ref.
\cite{ba2}
all tend to support the
largest
cross section of Ref. \cite{n}. The model calculation of Drachman and
Houston
\cite{d2}
denoted by a cross in Fig. 1 is consistent with  the experiment of Ref.
\cite{n1}.
This discrepancy has been partially resolved in the
study of Ref. \cite{ba2}, where it has been demonstrated that the present
model exchange potential is unique in being able to reproduce experimental
cross sections \cite{l1,l2} upto medium energies (about 60 eV) fairly
well.
Other theoretical  calculations are unable to reproduce \cite{bb,gh2,fk}
the experimental
trend of total cross section at different energies with a minimum around
5.1
eV.  The present elastic (full line) and momentum-transfer (dashed
line) static-exchange calculations  are consistent with the experiment of
Skalsey et al \cite{g2}.

\vskip -2.5cm
\postscript{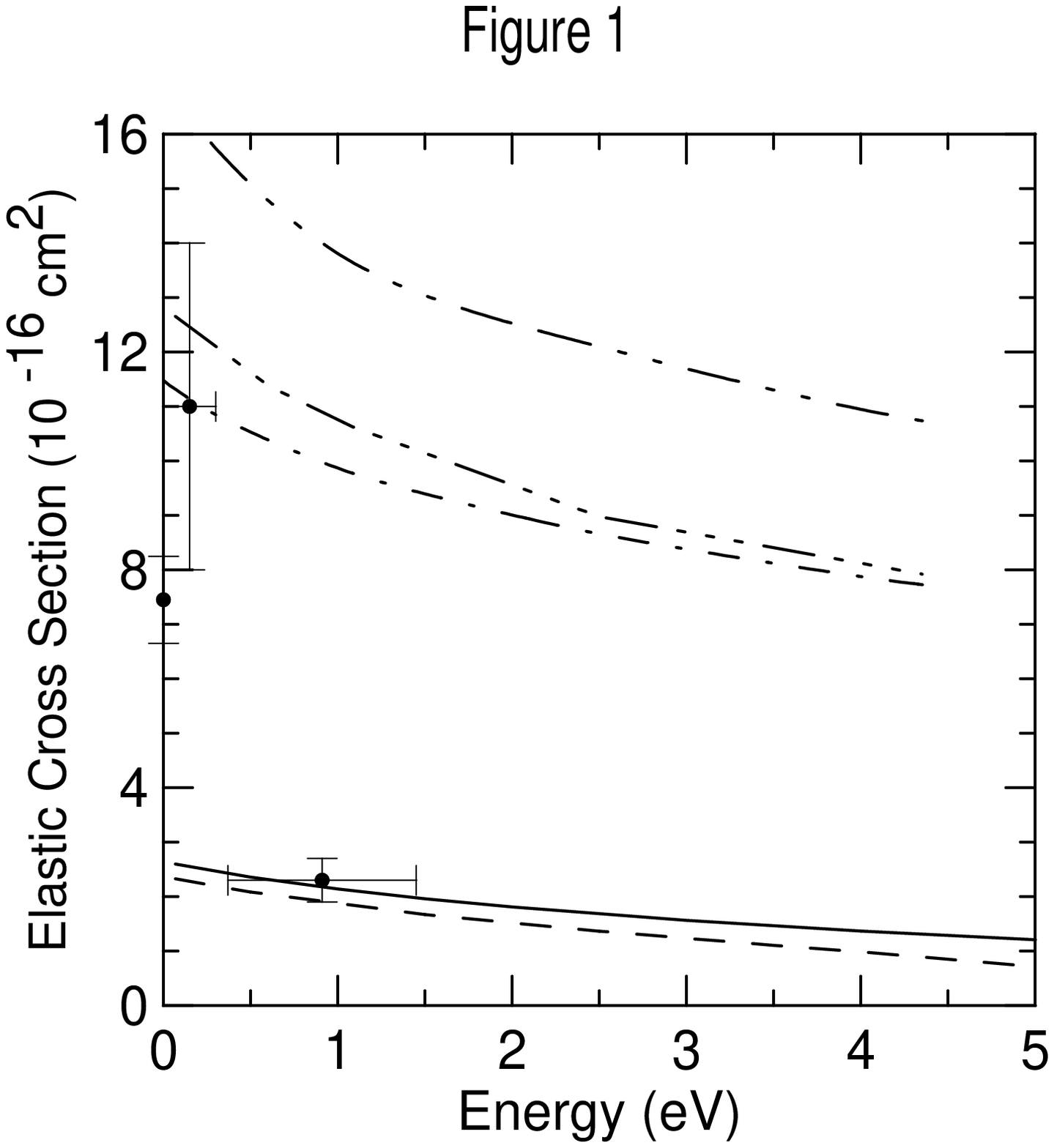}{1.0}    %**epsf**
\vskip -2.1cm

{ {\bf Fig. 1.}  Angle-integrated Ps-He  cross section at low positronium
energies:  present momentum transfer from static-exchange 
model (dashed line);  present elastic from 
static-exchange model (full line);
elastic from static-exchange of Ref. \cite{bb}  
(dashed-dotted line); of Ref. \cite{fk}
(dashed-double-dotted line); of Ref. \cite{gh2} 
 (dashed-triple-dotted line); theory of Ref. \cite{d2} (cross); 
experiments at 0 eV, 0.15 eV, and 0.9 eV of Refs. \cite{n1,n,g2},
respectively (solid circle).}

\vskip .2cm

Next we perform an S-wave calculation for the parameter
$^1Z_{\makebox{eff}}$.
In Fig. 2 we plot the results for the present
calculation of $^1Z_{\makebox{eff}}$ at different energies.
For comparison we also plot the results of previous calculation by
Barker and Bransden \cite{bb} and by Fraser and Kraidy \cite{fk} and the
existing three experimental data. Although Drachman and Houston \cite{d2}
did not calculate $^1Z_{\makebox{eff}}$ at different energies, their
low-energy value of 0.1 is in reasonable agreement with the present result
of 0.11 and
experiment of Refs. \cite{dh,dh1}. The
three
 experimental results with error bars cover the range of 0.07 to 0.31 for
 $^1Z_{\makebox{eff}}$.
Of the three experiments the one
by Duff and Heymann \cite{dh} with the   smallest error bar might be the
most accurate.

The much too small values of $^1Z_{\makebox{eff}}$ obtained in previous
calculations \cite{bb,fk} seem to be a consequence of a much stronger
(exchange) repulsion in the elastic channel of these models.  This is
reflected in the zero-energy cross section or the scattering length of
these calculations.
For a repulsive potential the low-energy cross section increases with
repulsion, consequently, the previous calculations  have led to unusually
large triplet scattering lengths  compared to the
present work. This is most clearly exhibited in a correlation
exhibited in Fig. 3 where we
plot $^1Z_{\makebox{eff}}$ versus triplet scattering length of different
calculations. The larger the scattering length the smaller is the
$^1Z_{\makebox{eff}}$. This correlation is similar to different
correlations observed in the study of Ps-H scattering in Ref. \cite{ba1a}.

\vskip 1.3cm

\vskip -3.9cm
\postscript{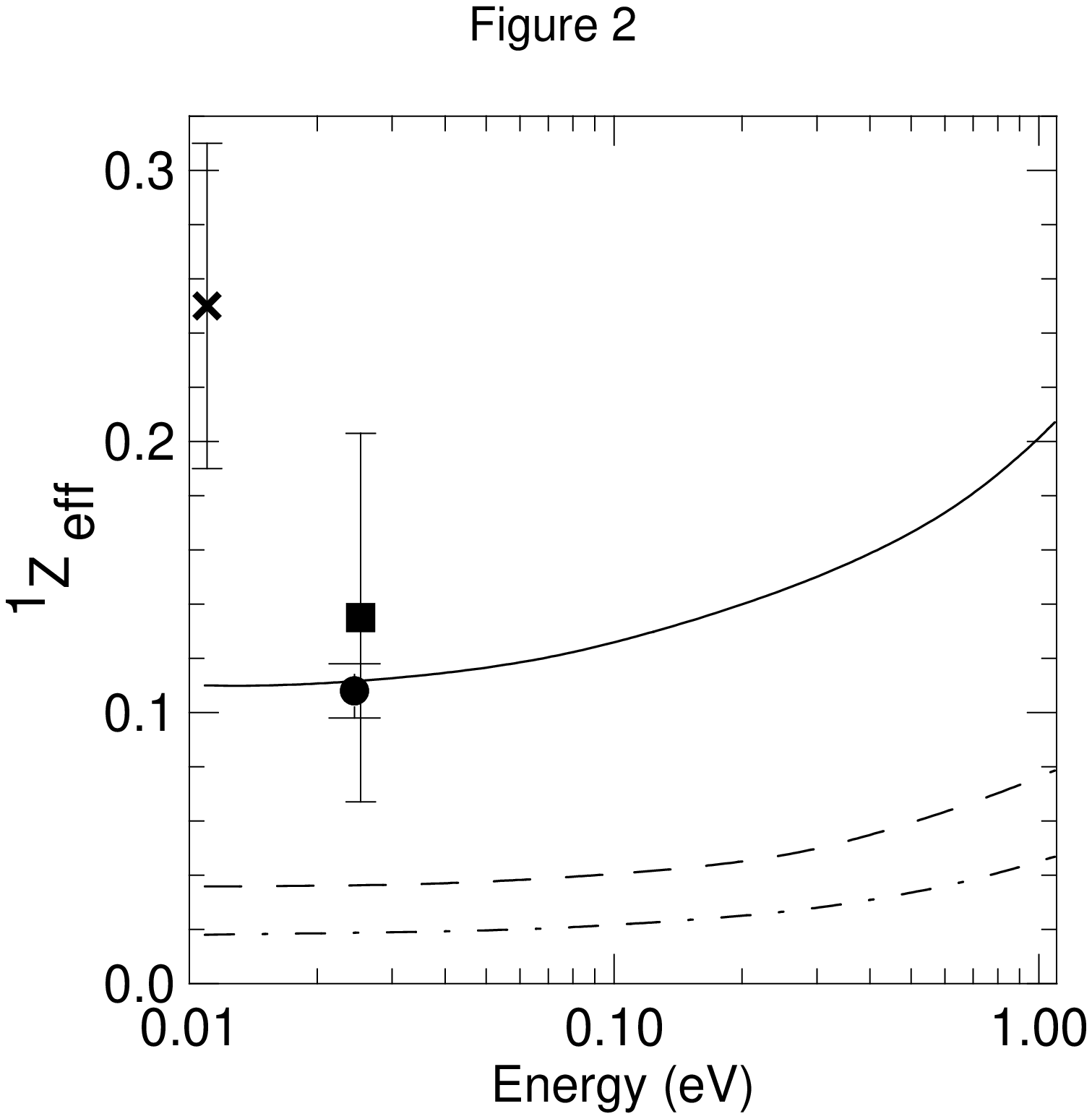}{1.0}    %**epsf**
\vskip -1.6cm

{ {\bf Fig. 2.}  The parameter $^1Z_{\makebox{eff}}$ at different
positronium  energies:
calculation including angular momenta $L=0,1,2 $ 
of Ref. \cite{fk} (dashed-dotted line),
of Ref. \cite{bb} 
(dashed line); calculation for $L=0$ of  present model 
(full line); the experimental points denoted solid circle, diamond, cross 
 taken from Refs.
\cite{dh,dh1,rk}, respectively.}

\vskip .2cm

We next comment on two aspects of the present calculation. First, we used
a two-term helium wave function. We also repeated our calculation with the
one-term helium wave function of Ref. \cite{fk1}  and the five-term wave
function used in Ref. \cite{ba2}. The results for  both the cross section
and $^1Z_{\makebox{eff}}$ suffer insignificant change with the change of
wave function. For $^1Z_{\makebox{eff}}$, the different results are
within the error bar of Ref. \cite{dh}; for cross section they  are also
within the error bar of Ref. \cite{g2}. Hence we do not believe the
present results to be so peculiar as to be of no general validity.
Secondly, we performed a $L=0$
calculation for $^1Z_{\makebox{eff}}$. At the experimental energies less
than 0.03 eV, the effect of higher partial waves is practically zero
(well within the error bar of Ref. \cite{dh});
at 1 eV this effect is quite small.

\vskip -3.cm

\postscript{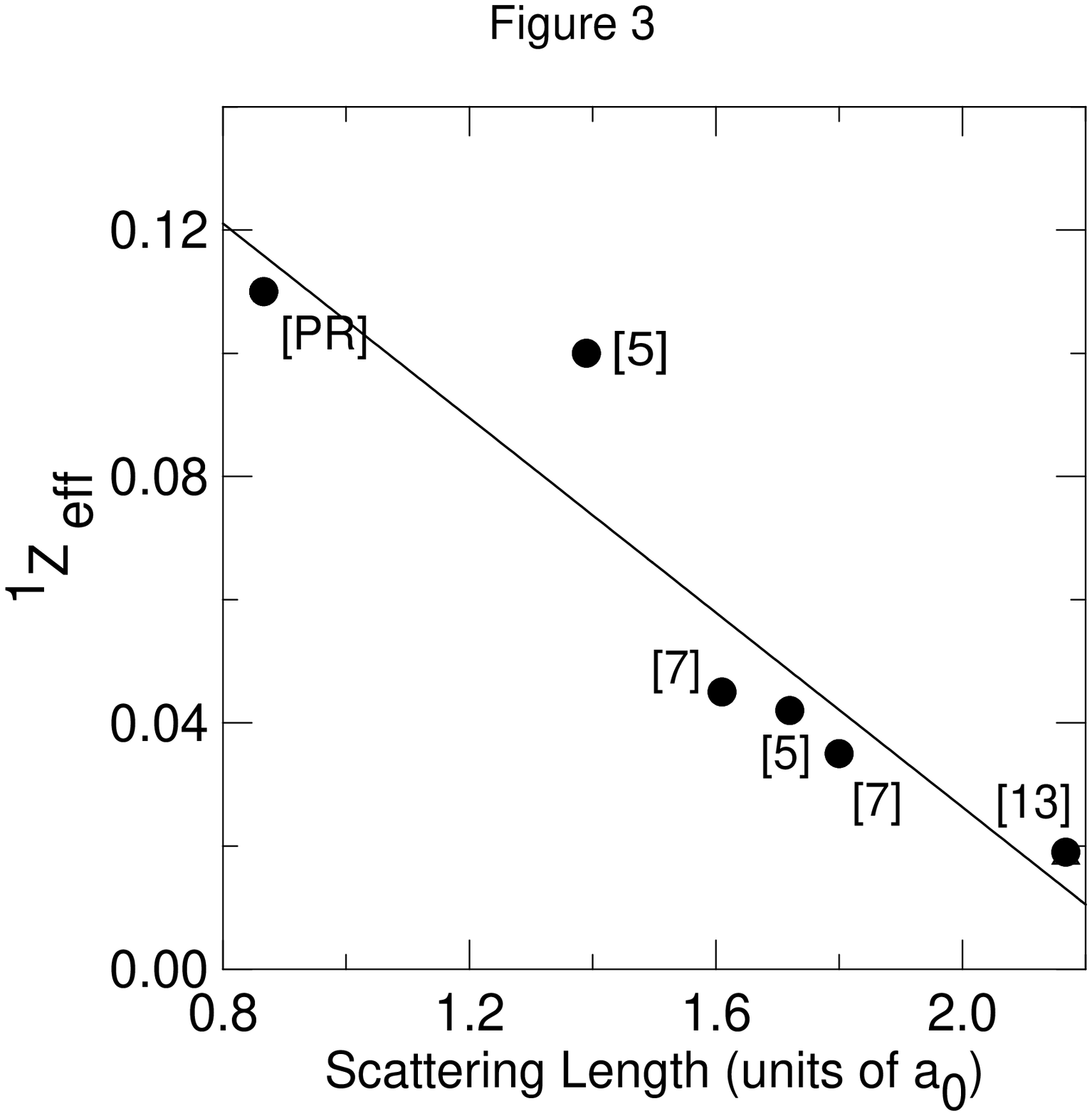}{0.8}    %**epsf**
\vskip -.4cm

{ {\bf Fig. 3.} The parameter $^1Z_{\makebox{eff}}$ versus scattering
length of various
calculations
 denoted by solid circle and labeled by reference number \cite{d2,bb,fk}.
The
present result
 is labeled by [PR], full line denotes a linear fit.}

\vskip .15cm

From the consideration above, we believe the present static-exchange
calculation as well as the previous calculations \cite{ba2} using the same
exchange potential provide a realistic account of very low-energy Ps-He
scattering. However, the precise agreement of the static-exchange
calculation with experiment is expected to be incidental. For a complete
understanding of this problem higher excited states of both Ps and He
should be incorporated in the model. The inclusion of Ps excitation
channels has been found \cite{ba2} to decrease the low-energy cross
sections and we might need to refit the low-energy cross sections by
changing the parameters $\alpha$ and/or $\beta$ of the potential in Eq.
(\ref{8}), as in Refs. \cite{ba1a,ba2}.

In conclusion, we have used a recently suggested nonlocal model exchange
potential \cite{ba1,ba1a,ba2,ba3} and applied it to the study of Ps-He
scattering at low energies. We have critically examined the
static-exchange calculation to see if it can account for satisfactorily
\cite{ba2} the measured cross sections of Refs. \cite{l1,g2} and the
`measured' value of the parameter $^1Z_{\makebox{eff}}$ of Refs.
\cite{dh,dh1}. The present calculation is in reasonable agreement with the
calculation of Ref. \cite{d2}. However, it is difficult to reconcile the
present calculation with the experiment of Ref.  \cite{rk}. In a previous
study \cite{ba2}, the present exchange potential has been found to
reproduce the low- to medium-energy cross sections of Refs. \cite{l1,g2}
well. This coupled with the present study seems to indicate that the
$^1Z_{\makebox{eff}}$ measurement of Ref.  \cite{dh} and the high- and
low-energy cross section measurements of Refs.
 \cite{l1} and \cite{g2} are consistent among each other as well as with
the present calculation which possibly provides a faithful description of
low-energy Ps-He scattering. We observe a correlation between
$^1Z_{\makebox{eff}}$ and triplet scattering length of various
calculations (Fig. 3), which
demonstrates that the smaller the scattering length the larger is the
value of $^1Z_{\makebox{eff}}$. This correlation
is similar to different correlations
observed between the low-energy Ps-H observables recently \cite{ba1a}.

We thank Prof. B. H. Bransden for suggesting this investigation, and for
his helpful comments and encouragements.
The work is supported in part by the CNPq and FAPESP
of Brazil.

%\vskip  2cm

\end{document}